\documentstyle[prl,aps]{revtex}
\draft
\begin{document}
\title{Hidden Integrability of a Kondo Impurity in an Unconventional
Host}
\author{Valery I. Rupasov}
\address{Department of Physics, University of Toronto, Toronto,
Ontario, Canada M5S 1A7\\
and Landau Institute for Theoretical Physics, Moscow, Russia}
\date{\today}
\maketitle
\begin{abstract}
We study a spin-$\frac{1}{2}$ Kondo impurity coupled to
an unconventional host in which the density of band states
vanishes either precisely at (``gapless'' systems) or
on some interval around the Fermi level (``gapped''systems).
Despite an essentially nonlinear band dispersion, the
system is proven to exhibit hidden integrability and is
diagonalized exactly by the Bethe ansatz.
\end{abstract}


In the Bethe ansatz (BA) approach to the theory of dilute
magnetic alloys \cite{TW,AFL}, initiated by Wiegmann
\cite{W1,W2} and Andrei \cite{A}, the conditions of (i)
a linear dispersion of band electrons near the Fermi
level and (ii) an energy independent electron-impurity
coupling play such a crucial role that up to recent time
they have been considered as necessary mathematical
conditions for integrability of impurity models.
However, it has recently been found \cite{R} that
integrability of the degenerate and $U\to\infty$
nondegenerate Anderson models is not destroyed by a
nonlinear dispersion of particles and an energy
dependent hybridization, but it becomes only hidden
\cite{RS}. The approach developed has allowed one
to study the ground state properties of a $U\to\infty$
Anderson impurity embedded in unconventional Fermi
systems, such as a BCS superconductor \cite{R2} and
a ``gapless'' host \cite{R3}. In the latter, a density
of band states is assumed to vanish precisely at the
Fermi level \cite{WF}.

In the BA approach, the spectrum of a metal host is
alternatively described in terms of interacting charge
and spin excitations rather than in terms of free particles
with spin ``up'' and ``down''. A choice of an appropriate
Bethe basis of a host is dictated by physical properties
of an impurity. In the Anderson model, spin and charge
excitations strongly interact, while in exchange models,
they are decoupled from each other. The latter stems
from a momentum independence of electron-impurity
scattering in exchange models. The Kondo model is thus
``simpler'' than the Anderson one, and its solution is
derived from that of the Anderson model in the limit
where an impurity energy lies much below the Fermi
level of a host.

However, in an unconventional host electron-impurity
scattering amplitudes even in exchange models are clear
to depend essentially on an electron momentum due to
an energy dependent density of band states. Therefore,
charge and spin degrees of freedom in exchange models
should be coupled to each other, that could lead to
novel features of the Kondo physics of unconventional
Fermi systems.

In this Letter, we report first results for the Kondo
model in an unconventional host. Despite an essentially
nonlinear dispersion of band electrons near the Fermi
level, the model is proven to exhibit hidden
integrability. As in the unconventional Anderson model
\cite{R}, an auxiliary $\tau$-space related to the
particle energy is introduced to show that the multiparticle
scattering process is factorized into two-particle ones.
Multiparticle functions in an auxiliary $x$-space related
to the particle momentum are then found as a result of
an integral ``dressing'' procedure of the BA functions
in the $\tau$-space. Thus, factorizability of scattering
in the $x$-space is hidden and becomes visible only in
the limit of large interparticle separations.

Integrability of the unconventional Kondo model is clear
to provide us a powerful theoretical tool in studies of
the Kondo physics in such systems of fundamental interest
as superconductors and unconventional Fermi systems, where
a band particle dispersion cannot be linearized near
the Fermi level.

We start with the standard s-d exchange (Kondo)
model with an arbitrary band electron dispersion
$\epsilon(k)$. An effective 1D Hamiltonian of the
model is written in terms of the Fermi operators
$c^\dagger_\sigma(\epsilon)$ [$c_\sigma(\epsilon)$]
which create (annihilate) an electron with a spin
$\sigma=\uparrow,\downarrow$ in an $s$-wave state
of energy $\epsilon$,
\begin{equation}
H=\sum_{\sigma}\int_{C}\frac{d\epsilon}{2\pi}\epsilon
c^\dagger_\sigma(\epsilon)c_\sigma(\epsilon)
+\sum_{\sigma,\sigma'}
\int_{C}\frac{d\epsilon}{2\pi}
\frac{d\epsilon'}{2\pi}I(\epsilon,\epsilon')
c^\dagger_\sigma(\epsilon)
\left(\vec{\sigma}_{\sigma\sigma'}\cdot\vec{S}\right)
c_{\sigma'}(\epsilon').
\end{equation}
Here, $\vec{\sigma}$ are the Pauli matrices, $\vec{S}$ is
the impurity spin operator. An effective electron-impurity coupling
$I(\epsilon,\epsilon')=\frac{1}{2}I\sqrt{\rho(\epsilon)\rho(\epsilon')}$
combines the exchange coupling constant $I$ and the density
of band states $\rho(\epsilon)=dk/d\epsilon$. The integration
contour $C$ contains two intervals, $C=(-D,-\Delta)\oplus(\Delta,D)$,
where $D$ is the half band width, and $2\Delta$ is the energy
gap in a gapped host, while in a gapless host $\Delta=0$.
In what follows, we restrict our consideration to the case
of $S=\frac{1}{2}$. The electron energies and momenta in Eq.
(1) and hereafter are taken relative to the Fermi values,
which are set to be equal to zero.

We look for one-particle eigenstates of the system
in the form
\begin{equation}
|\Psi_1>=\sum_{\sigma}\sum_{s=0,1}\int_{C}\frac{d\epsilon}{2\pi}
\psi_{\sigma;s}(\epsilon)c^\dagger_\sigma(\epsilon)
\left(S^+\right)^s|0>,
\end{equation}
where the vacuum state $|0>$ contains no electrons
and the impurity spin is ``down''. The Schr\"odinger
equation for the auxiliary wave function
$\phi_{\sigma;s}(\epsilon)=\psi_{\sigma;s}(\epsilon)/\sqrt{\rho(\epsilon)}$
is easily found to be
\begin{mathletters}
\begin{eqnarray}
&&(\epsilon-\omega)\phi_{\sigma;s}(\epsilon|\omega)+
\frac{1}{2}I
\left(\vec{\sigma}_{\sigma\sigma'}\cdot\vec{S}_{ss'}\right)
A_{\sigma';s'}(\omega)=0\\
&&A_{\sigma;s}(\omega)=\int_{C}\frac{d\epsilon}{2\pi}
\rho(\epsilon)\phi_{\sigma;s}(\epsilon|\omega),
\end{eqnarray}
where $\omega$ is the eigenenergy.

To simplify notation we omit hereafter the spin indexes.
Inserting the general solution of Eq. (4a),
\end{mathletters}
\begin{mathletters}
\begin{equation}
\phi(\epsilon|\omega)=
2\pi\delta(\epsilon-\omega)\chi-
\frac{\frac{1}{2}I}
{\epsilon-\omega-i0}\left(\vec{\sigma}\cdot\vec{S}\right)
A(\omega),
\end{equation}
with an arbitrary spinor $\chi$, into Eq. (4b), one
obtains
\begin{equation}
\left[1+\frac{1}{2}I\Sigma(\omega)
\left(\vec{\sigma}\cdot\vec{S}\right)\right]A(\omega)=
\rho(\omega)\chi.
\end{equation}
Here, the self-energy $\Sigma(\omega)$ is found to be
\end{mathletters}
\begin{equation}
\Sigma(\omega)=\int_{C}\frac{d\epsilon}{2\pi}
\frac{\rho(\epsilon)}{\epsilon-\omega-i0}=
P\int_{C}\frac{d\epsilon}{2\pi}
\frac{\rho(\epsilon)}{\epsilon-\omega}+\frac{i}{2}\rho(\omega)
=\Sigma'(\omega)+i\Sigma''(\omega),
\end{equation}
where $P$ stands for the principal part.

It is convenient to rewrite Eq. (3a) for the Fourier
image of the function $\phi(\epsilon|\omega)$,
\begin{equation}
\phi(\tau|\omega)=\int_{-\infty}^{\infty}\frac{d\epsilon}{2\pi}
\phi(\epsilon|\omega)\exp{(i\epsilon\tau)}.
\end{equation}
Then, the equation (4a) takes the form
\begin{equation}
\left(-i\frac{d}{d\tau}-\omega\right)\phi(\tau|\omega)
+\frac{1}{2}I\delta(\tau)\left(\vec{\sigma}\cdot\vec{S}\right)A=0.
\end{equation}
Its solution is easily found to be
\begin{equation}
\phi(\tau|\omega)=e^{i\omega\tau}\left\{
\begin{array}{rl}
\chi,&\tau<0\\
{\bf R}(\omega)\chi,&\tau>0
\end{array}
\right.
\end{equation}
with the electron-impurity scattering matrix
\begin{mathletters}
\begin{equation}
{\bf R}(\omega)=u(\omega)+2v(\omega)\left(\vec{\sigma}\cdot\vec{S}\right),
\end{equation}
where the parameters $u(\omega)$ and $v(\omega)$ are determined
by the equations
\begin{eqnarray}
u+v&=&1-i\frac{\frac{1}{4}I\rho(\omega)}{1+\frac{1}{4}I\Sigma(\omega)}\\
u - 3v&=&1+i\frac{\frac{3}{4}I\rho(\omega)}{1-\frac{3}{4}I\Sigma(\omega)}.
\end{eqnarray}

In a metal, where an electron-impurity scattering is
energy independent, one may introduce the permutation
operator of electron spins,
${\bf P}=\delta_{\sigma_1,\sigma'_2}\delta_{\sigma_2,\sigma'_1}$
as an electron-electron scattering matrix to factorize
multiparticle scattering and to construct thus Bethe ansatz
eigenfunctions of the system. In an unconventional host,
multiparticle scattering is not factorized by the
permutation operator because of an energy dependence
of electron-impurity scattering amplitudes.

Let us introduce a two-particle scattering matrix of
electrons with the energies $\omega_1$ and $\omega_2$,
\end{mathletters}
\begin{mathletters}
\begin{equation}
\phi(\tau_1>\tau_2|\omega_1,\omega_2)=
{\bf r}_{12}(\omega_1,\omega_2)\phi(\tau_1<\tau_2|\omega_1,\omega_2),
\end{equation}
by the most general $SU(2)$-symmetric expression
\begin{equation}
{\bf r}(\omega_1,\omega_2)=
\frac{h(\omega_1)-h(\omega_2)-i{\bf P}}
{h(\omega_1)-h(\omega_2)-i}.
\end{equation}
Then multiparticle scattering in the system of host
electrons is well known to be factorized at an arbitrary
function $h(\omega)$, since the matrix ${\bf r}_{ij}$,
where the particle index $j=1,2,3$ is assumed to incorporate
also its energy, obeys the Yang-Baxter factorization conditions
\cite{TW,AFL}
\end{mathletters}
\begin{mathletters}
\begin{equation}
{\bf r}_{12}{\bf r}_{13}{\bf r}_{23}=
{\bf r}_{23}{\bf r}_{13}{\bf r}_{12}.
\end{equation}
Introducing an impurity fixes an expression for the function
$h(\omega)$. To factorize multiparticle scattering in the
presence of the impurity with ${\bf R}$ matrix derived in
Eq. (10), one needs to solve the Yang-Baxter equations
involving two electrons and the impurity,
\begin{equation}
{\bf r}_{12}{\bf R}_{10}{\bf R}_{20}=
{\bf R}_{20}{\bf R}_{10}{\bf r}_{12},
\end{equation}
that gives
\begin{equation}
h(\omega)=\frac{u(\omega)}{v(\omega)}.
\end{equation}

Thus, the multiparticle scattering in the unconventional
Kondo model is proven to be factorized in the auxiliary
$\tau$-space related to the particle energy, and the
$N$-particle eigenfunctions of the model,
$\Phi(\tau_1,\ldots,\tau_N)$ are written in the standard
BA form. The $N$-particle eigenfunctions in the auxiliary
$x$-space related to the particle momentum, $\Psi(x_1,\ldots,x_N)$,
are found as a result of a ``dressing'' procedure
\end{mathletters}
\begin{mathletters}
\begin{equation}
\Psi(x_1,\ldots,x_N)=\int_{-\infty}^{\infty}
\Phi(\tau_1,\ldots,\tau_N)\prod_{j=1}^{N}u(x_j|\tau_j)d\tau_j,
\end{equation}
with the dressing function
\begin{equation}
u(x|\tau)=\int_{C}\frac{d\epsilon}{2\pi}
\rho(\epsilon)e^{i[k(\epsilon)x-\epsilon\tau]},
\end{equation}
where $k(\epsilon)$ is the inverse dispersion. In a metal
host, where $\epsilon(k)=k$, the dressing function is nothing
but the Dirac delta function, $u(x|\tau)=\delta(x-\tau)$,
and hence the $x$ and $\tau$ representations coincide.

Imposing periodic boundary conditions on the wave function
$\Psi(x_1,\ldots,x_N)$ on the interval of size $L$ leads
in the standard manner \cite{TW,AFL} to the BA
equations
\end{mathletters}
\begin{mathletters}
\begin{eqnarray}
\exp{(ik_jL)}\varphi_c(\omega_j)&=&
\prod_{\alpha=1}^{M}\frac{h_j-\lambda_\alpha-\frac{i}{2}}
{h_j-\lambda_\alpha+\frac{i}{2}}\\
\varphi_s(\lambda_\alpha)
\prod_{j=1}^{N}\frac{\lambda_\alpha-h_j-\frac{i}{2}}
{\lambda_\alpha-h_j+\frac{i}{2}}
&=&-\prod_{\beta=1}^{M}
\frac{\lambda_\alpha-\lambda_\beta-i}{\lambda_\alpha-\lambda_\beta+i}
\end{eqnarray}
where $M$ is the number of particles with spin ``down'',
$k_j\equiv k(\omega_j)$, and $h_j\equiv h(\omega_j)$.
The eigenenergy $E$ and the $z$ component of the total
spin of the system $S^z$ are found to be
\begin{equation}
E=\sum_{j=1}^{N}\omega_j,\;\;\;S^z=\frac{1}{2}+\frac{N}{2}-M.
\end{equation}
In Eqs. (13) the phase factors
\end{mathletters}
\begin{mathletters}
\begin{eqnarray}
\varphi_c(\omega)&=&
\frac{1+\frac{1}{4}I\Sigma'(\omega)-\frac{i}{2}\frac{1}{4}I\rho(\omega)}
{1+\frac{1}{4}I\Sigma'(\omega)+\frac{i}{2}\frac{1}{4}I\rho(\omega)}\\
\varphi_s(\lambda)&=&\frac{\lambda-\frac{i}{2}}{\lambda+\frac{i}{2}}
\end{eqnarray}
describe the scattering of charge and spin excitation
of the host on the impurity. The equations (13) solve
exactly the problem of diagonalization of the s-d
exchange (Kondo) model in an unconventional host.

It should be emphasized that these equations are incomplete
yet in the case of a gapped host, since they do not account
for electron-impurity bound states with eigenenergies lying
inside the gap. However, for a further analysis of the problem
one needs to specify physical characteristic of a host,
therefore we are going to address the thermodynamic properties
of a Kondo impurity in unconventional Fermi systems in
separated publications.

\end{mathletters}

\end{document}